\documentclass{aastex}                              

\usepackage{emulateapj5}

\newcommand{\sigs}{$\sigma_{*}$}
\newcommand{\sige}{$\sigma_{e}$}
\newcommand{\Mb}{$M_{bul}$}
\newcommand{\Mbh}{$M_{\bullet}$}
\newcommand{\Msol}{$M_{\odot}$}
\newcommand{\asecx}{$^{\prime \prime} \!\!$}

\newcommand{\Oiii}{[OIII] $\lambda $5007 }
\newcommand{\fwoiii}{$\rm FWHM_{[OIII]}$ }

\newcommand{\magx}{$^{\rm m} \!\!$}
\newcommand{\Mgb}{Mg$ b$}
\newcommand{\Cat}{Ca~II triplet}

\begin{document}

\title{Black Hole Mass, Velocity Dispersion
and the Radio Source in AGN}

\author{Charles H. Nelson}

\affil{Physics Dept., University of Nevada, Las Vegas, Box 4002, 
4505 Maryland Pkwy., Las Vegas, NV 89154, cnelson@physics.unlv.edu}

\begin{abstract}

The recent discovery of a correlation between nuclear black hole mass,
\Mbh, and the stellar velocity dispersion, \sigs
\citep{geb2000a,fer2000}, in elliptical galaxies and spiral bulges,
has raised the question whether such a relationship exists for
AGN. Estimates of \Mbh \ for many AGN, made using reverberation
mapping techniques, allow exploration of the relationship between
black hole mass, the host galaxy and the energetics of nuclear
emission.  However, since only a few AGN have both \Mbh \ and \sigs \
measurements, we use the [OIII] $\lambda $ 5007 emission line widths
on the assumption that for most AGN the forbidden line kinematics are
dominated by virial motion in the host galaxy bulge. We find that a
relation does exist between \Mbh \ and [OIII] line width for AGN which
is similar to the one found by \citet{geb2000a}, although with more
scatter as expected if secondary influences on the gas kinematics are
also present.  Our conclusion is that both active and inactive
galaxies follow the same relationship between black hole mass and
bulge gravitational potential.  We find no compelling evidence for
systematic differences in the mass estimates from reverberation
mapping and stellar dynamics.  We also find that for radio quiet AGN
the radio power and black hole mass are highly correlated linking
emission on scales of kiloparsecs with the nuclear energy source.

\end{abstract}


\section{Introduction}

In the standard model for an active galactic nucleus (AGN), radiant
energy is released through accretion onto a supermassive ($10^6 -
10^9$ \Msol) black hole. However, until recently, there have been few
mass estimates for nuclear black holes in AGN. Kinematic studies of
nuclear gas disks \citep[{\it e.g.} ][]{harms94} and stellar dynamical
studies, using high spatial resolution ground-based spectroscopy
\citep[{\it e.g.} ][]{kor88} and more recently with the Space
Telescope Imaging Spectrograph (STIS) \citep[{\it e.g.}
][]{bower2000}, have produced numerous central mass estimates mostly
for normal galaxies.  However, in application to AGN these techniques
are severely limited.  Since roughly one galaxy in a hundred has a
luminous active nucleus even the nearest examples are relatively
distant reducing the effective spatial resolution.  Furthermore,
stellar dynamical techniques require high signal-to-noise measurements
of stellar absorption features which are often lost in the glare of a
bright active nucleus. Also emission lines from ionized gas at small
radii can be influenced by relativistic jets or winds associated with
the AGN.

These difficulties can be overcome, however, using reverberation
mapping techniques \citep[{\it e.g.} ][]{np97}. Monitoring the
variability of the continuum and emission line fluxes in broad line
AGN, shows that the two are highly correlated but that the emission
lines lag behind the continuum by times ranging from days to months.
The lag is best explained as a light travel time effect and can
be used to estimate the size of the Broad Line Region (BLR). If one
further assumes that the BLR kinematics are largely Keplerian,
one can use the line widths as a characteristic velocity and
in turn determine the total mass contained within the BLR. Strong
evidence for Keplerian motion exists for at least one well-studied
Seyfert, NGC 5548. Results for several different emission lines, each
with different lags and therefore spanning a range of distances from
the nucleus, give consistent values for the central mass
\citep{pw99}. Reverberation studies have now yielded \Mbh \ estimates
for numerous Seyfert 1 galaxies and quasars \citep[{\it e.g.}
][]{kaspi2000}. However, there have been some claims that the masses
determined from these variability studies are systematically low
\citep{wandel99,ho99}.

Recent work has focused on the distributions of black hole properties
and the relationship to their host galaxies.  \citet{korrich} first
found a correlation between \Mbh \ and the mass of the spheroidal
component, \Mb. In a comprehensive study using ground-based
spectroscopy and HST imaging, \citet{mag98}, presented strong evidence
in support of the the \Mbh \ -- \Mb \ relation.  More recently, two
studies \citep{geb2000a,fer2000} reported relationships between \Mbh \
and \sige, the stellar velocity dispersion obtained within a large
aperture extending to the galaxy effective radius, thus insensitive to
the influence of the black hole.  The correlation between these two
parameters is remarkably strong suggesting a link between the
formation of the bulge and the development of the black hole.

In this paper we combine reverberation mapping measurements of \Mbh \
in AGN with Narrow Line Region (NLR) gas and bulge stellar kinematic
measurements corresponding to motion well beyond influence of the
black hole to compare the nuclei of active and normal galaxies.  We
also investigate the possibility of a correlation between black hole
mass and the radio luminosity in AGN.  Section 2 describes the data
collected from the literature. Section 3 presents
the \Mbh -- \sigs \ relation for AGN. In section 4 we consider the
relationship between \Mbh \ and the radio luminosity and in Section 5
we summarize our conclusions.

\section{Data}

We start with all AGN with black hole masses, measured by the
reverberation mapping technique: 17 Seyfert 1 galaxies \citep[as
compiled by ][]{wpm99} and 17 PG quasars \citep{kaspi2000}.  For most
of the galaxies, we use the black hole masses tabulated by
\citet{kaspi2000} evaluated using the mean H$\beta$ profile width.
\Mbh \ values for three galaxies (Mkn 279, NGC 3516, and NGC 4593)
were taken from \citet{ho99}. For as many objects as possible we
obtained published measurements of \sigs, \fwoiii, and the radio flux.
The adopted values are presented in Table \ref{tab:agnstuff}.  For the
Seyferts, \fwoiii \ is taken from \citet{w92a}.  For the quasars
\fwoiii \ values were only adopted if the spectroscopy was of medium
to high resolution ($R > 1500$).  \sigs \ in AGN host galaxies has
been determined using spectral regions around both the \Cat \ and
\Mgb.  The values used here have been published in \citet{smith90},
\citet{tdt} and \citet{nw95} \citep[see ][for references on individual
galaxies]{nw95}.  Radio luminosities are from \citet{w92a} for
Seyferts and from \citet{Ke89} for PG quasars. We have scaled
observations at 5Ghz assuming a power-law index of 0.7 and assume $\rm
H_0=80 km s^{-1} Mpc^{-1}$. We also calculate $R$, the ratio of the
radio to optical flux, using the mean continuum
luminosities from \citet{kaspi2000}.

We point out that the stellar and gas kinematic measurements are from
small aperture observations typically covering $2-3$\asecx, somewhat
smaller than the apertures used to measure \sige.  Nevertheless, we
can be confident that these measurements are unaffected by the
presence of the black hole. First, since the AGN are at larger
distances than the normal galaxies for which \Mbh \ has been
determined, the aperture typically corresponds to physical scales of a
kpc or more, actually comparable to the bulge effective radius. Second
\citet{nw96} found that Seyferts were offset from the Faber-Jackson
relation ($M_{bul} \propto \rm log \sigma_*$) for normal galaxies in
the sense of having lower velocity dispersions than normal galaxies of
the same \Mb. Their interpretation was that Seyfert bulges have lower
$M/L$ ratios than normal spirals, exactly the opposite of what one
would expect if the stellar kinematics were strongly influenced by a
massive nuclear black hole.

\section{The \Mbh -- \sigs \ Relation for AGN}

Recently, \citet{geb2000b} included Seyferts in their plot of \sigs \
vs. \Mbh \ (in some cases choosing different values for \sigs \ than in
Table \ref{tab:agnstuff}) and conclude that AGN show no significant
difference from the overall relation.  Unfortunately, \sigs \ values
are available in the literature for only a hand-full of the AGN with
reverberation mapping estimates for \Mbh.  Therefore a definitive
comparison of AGN and normal galaxies is not possible with these data.
However, \citet{nw96} have shown that for the majority of Seyfert
galaxies, a moderately strong correlation between \fwoiii \ and \sigs
\ exists indicating roughly equal absorption and emission line
widths. Thus the \Oiii \ profiles are dominated by virial motion in
the bulge potential. As might be expected, a fair amount of real ({\it
i.e.}  not due to measurement error) scatter does exist.  \citet{nw96}
used the deviation from purely virial gas motion to investigate
possible secondary influences on the NLR kinematics. Their results
confirmed conclusions from previous studies \citep[{\it e.g.}
][]{w92b,w92c} that the interaction of NLR gas with a relatively
strong kiloparsec-scale linear radio source can produce non-virial gas
acceleration. They also found a weak tendency for interacting systems
and mergers to have broader emission lines than expected from purely
virial motion, an issue that may be important in quasars which are
more likely to be interacting systems \citep{mcleod94}. Thus, keeping
the issue of secondary influences on the NLR kinematics in mind, we
can proceed with the idea that the primary influence on the forbidden
line widths in AGN is the bulge potential.

\hspace*{-26pt}
\scalebox{0.7}{\includegraphics[130,72][490,670]{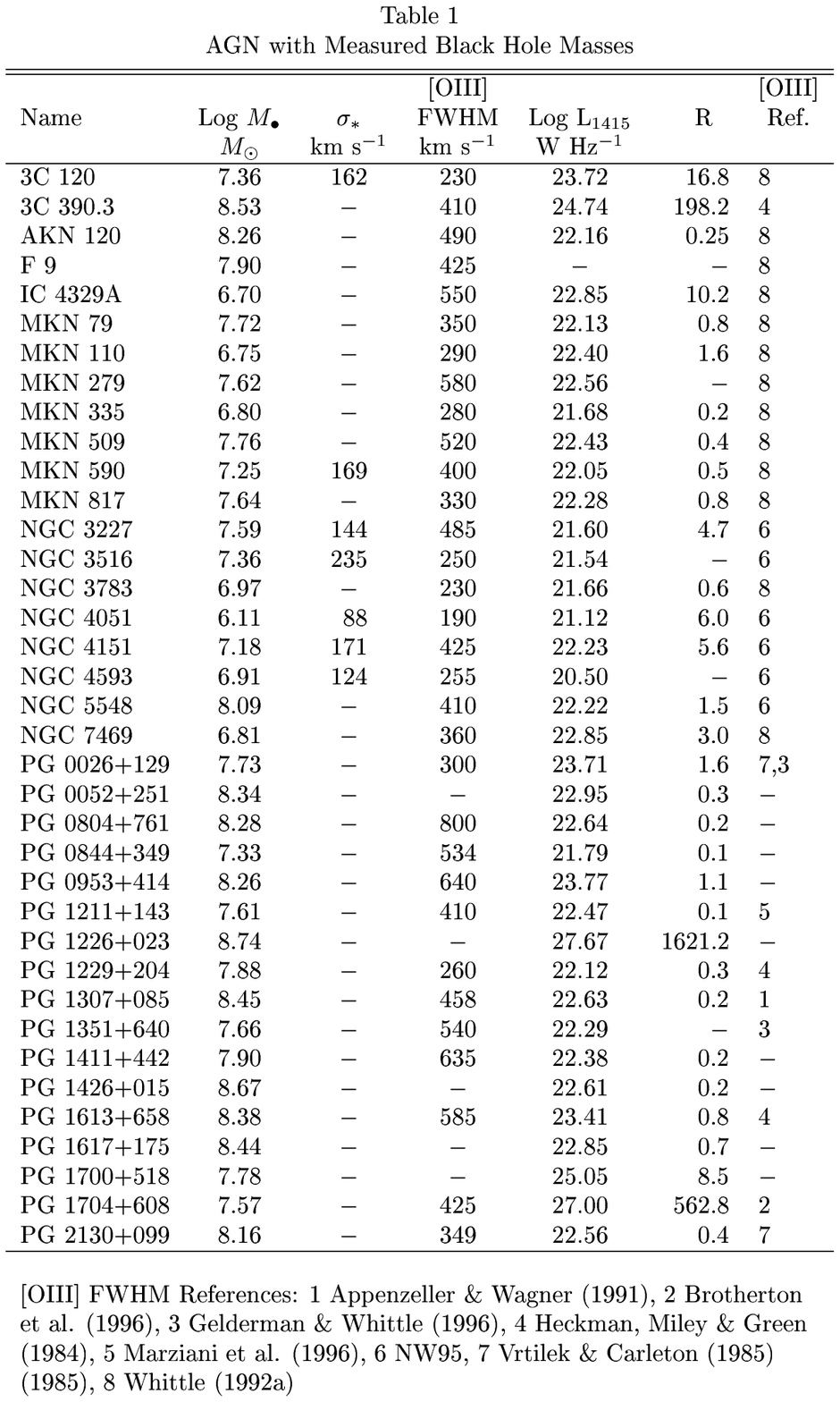}}
\label{tab:agnstuff}

In Figure \ref{fig:bhsig} we plot \fwoiii/2.35 \ or \sigs \ vs. \Mbh.
Filled symbols show the \fwoiii values \ for the galaxies in Table
\ref{tab:agnstuff}, $\bullet$ for Seyferts, \protect
\rule[0pt]{5pt}{5pt} for the PG quasars, $+$ signs show AGN for
which \sigs \ measurements exist and triangles show \Mbh \ and
\sige \ from \citet{geb2000a}.  The similarity in the trends between
the data from Table 1 and the \citet{geb2000a} sample is striking.  The
long solid line is the relation determined by \citet{geb2000a} and the
shorter dashed line is our fit to the AGN, using the 
ordinary least-squares bisector \citep{isobe} which gives good results for
large uncertainties in both variables. The correlation for the AGN is
moderately strong ($R=0.51 $, $P(null)=0.5$\%) with a slope
$3.7\pm0.7$ and intercept $-0.5\pm0.1$.  The AGN relation is
slightly shifted to larger widths for a given \Mbh \ than the
\citet{geb2000a} relation and is more scattered. Also,
there seems to be no statistical difference between the distributions
of Seyferts and quasars.
\hspace*{-28pt}
\rotatebox{-90}{\scalebox{0.36}{\includegraphics{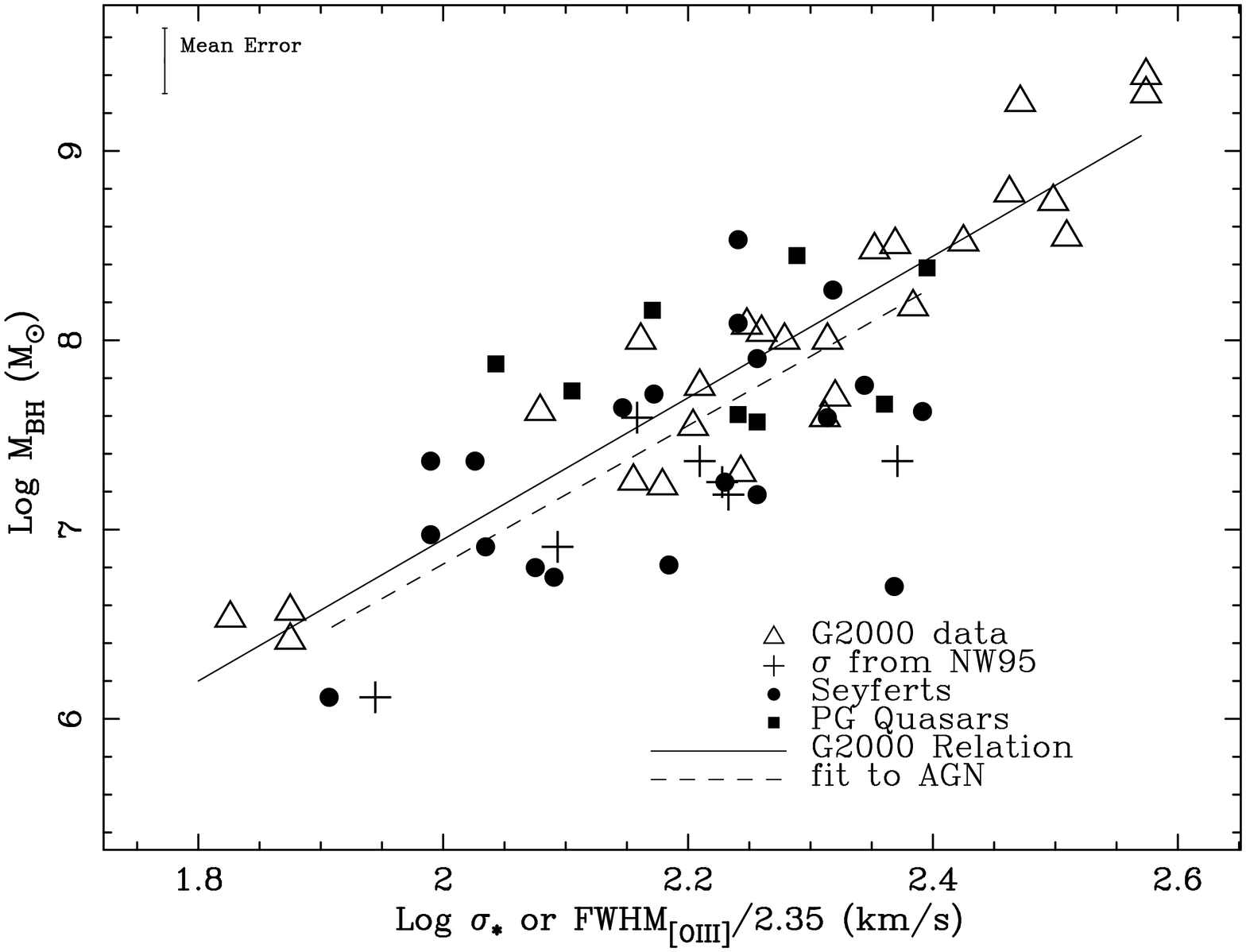}}}
\figcaption{
The data from \citet{geb2000a} is shown as $\triangle$ and the
solid line is the \Mbh \ -- \sige \ relation derived there. Objects
with \sigs \ from \citet{nw96} are shown as $+$. Objects for which
the [OIII] line widths have been converted to $\sigma$ are shown
as $\bullet$ for Seyferts and as \protect \rule[0pt]{4pt}{4pt}
and the dashed line is the fit.
\label{fig:bhsig}}
\medskip

The slope on the fit agrees well with that found by 
\citet{geb2000a} who derived a different value than \citet{fer2000}
(3.75 as opposed to 4.8). \citet{geb2000a} discuss various explanations for
the difference in the two values. Although our result 
supports a shallower slope, an analysis using stellar velocity dispersion
instead of emission line width would be more conclusive.

We can be more quantitative about the similarity of the relation for
elliptical galaxies and that for the AGN, by considering the origin of
the scatter. The mean $1-\sigma$ error bar in log \Mbh, plotted in the
upper left, is $+0.17/-0.20$ while the r.m.s. deviation of the filled
symbols in Figure \ref{fig:bhsig}, relative to the \citet{geb2000a}
relation is 0.55.  Since we expect that emission line widths can be
determined to accuracies of $\sim$ 10\% for the spectral resolution of
the observations, we expect that a large component of the scatter is
not due to measurement error but reflects real deviations perhaps
indicating secondary influences on NLR kinematics. If we calculate the
difference from the \citet{geb2000a} relation, this time in log
FWHM$_{\rm [OIII]}$, we find a small offset, $\Delta \rm
log~FWHM_{[OIII]} = 0.06$, with a distribution of r.m.s. width
0.14. \citet{nw96} used differences in emission and absorption line
widths, $\Delta W \equiv \rm log FWHM_{[OIII]} - log 2.35 \sigma_*$,
to investigate non-virial influences on emission line kinematics. For
their entire Seyfert sample they found $\Delta W = 0.0$ and an
r.m.s. width of 0.2. Excluding interacting galaxies and objects with
relatively luminous linear radio sources they found $\Delta W = -0.1$,
thus slightly narrower emission lines than absorption lines, and an
r.m.s. width of 0.13.  Thus it is reasonable that the scatter and the
small offset indicate non-virial contributions to the line widths in
some objects. The fact that the r.m.s. deviation is comparable to that
found in \citet{nw96} is consistent with this interpretation.
Unfortunately, the current sample includes no Seyferts with strong
linear radio sources and, due to their larger distances, the radio
maps for the quasars are mostly unresolved on kpc scales \citep[see
{\it e.g.}][]{Ke94,Ku98}.  Therefore, no strong statements regarding
non-virial gas kinematics can be made.

The relationship between \Mbh \ and the \fwoiii \ in AGN is precisely
what we would expect if AGN follow the \Mbh \ -- \sigs \ relation and
if the primary influence on NLR kinematics is the bulge gravitational
potential.  Although one might point out that the AGN plotted with
\sigs \ do seem to lie systematically below the relation, the reality
of such a shift based on only 7 objects is not compelling.  Our
primary conclusion is that the relationship between the mass of the
central dark object and the depth of the gravitational potential in
AGN is the same as for normal galaxies.  Clearly more velocity dispersion
measurements for AGN are needed. However, in fainter objects 
detailed studies of the  [OIII] emission line profiles 
accounting for non-virial kinematic components could provide a valuable
substitute.

\hspace*{-26pt}
\rotatebox{-90}{\scalebox{0.36}{\includegraphics{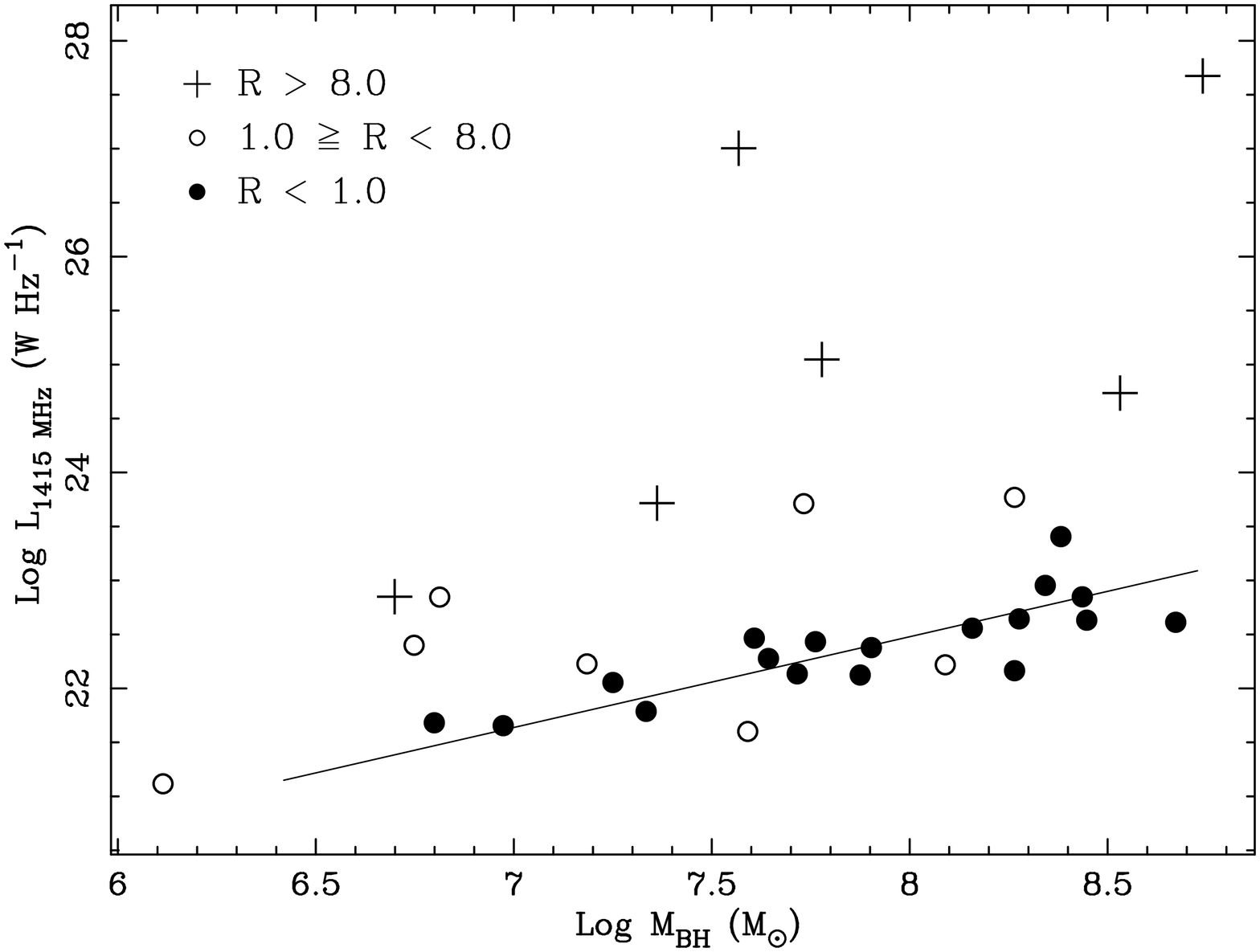}}}
\figcaption{Log L$_{1415 MHz}$ \ vs. \Mbh \ for AGN. 
\label{fig:bhsigrad}}
\medskip

These results seem to conflict with previous reports that AGN have
lower black hole to bulge mass ratios than normal galaxies
\citep{wandel99} or alternatively that the reverberation mapping
estimates of \Mbh \ are systematically low \citep{ho99}.  
Figure \ref{fig:bhsig} suggests that neither of these is
correct and that the differences must lie in the values of
\Mb.  A number of factors may have contributed
to this apparent disagreement.  First, the \Mbh \ -- \sige \ relation
derived by \citet{geb2000a} has been evaluated using only stellar
dynamical black hole mass estimates from three-integral modeling.
The result is that the black hole masses are about a factor of 3 lower
than the ones published by \citet{mag98} which were used in
\citet{wandel99}.  Second \citet{nw96} found that Seyferts are offset
from the Faber-Jackson relation for normal galaxies having brighter
bulges by on average about 0\magx .6 at the same velocity
dispersion. More generally bulge-disk decomposition is a difficult
task even for normal, nearby galaxies. The problem is compounded in
Seyfert 1 galaxies by their smaller apparent bulge sizes,
due to larger mean distances, and the bright point source
component.  Thus, much of the discrepancy can be resolved by
acknowledging that as much or greater uncertainty exists in the bulge
luminosity of Seyferts as exists in the estimates of \Mbh \ by any of
the available techniques.

\section{Radio Luminosity and Black Hole Mass in AGN}

Using reliable measurements of black hole mass in AGN we can compare
parameters related to the energetics of AGN with the size of the
central engine. For example \citet{kaspi2000} found a strong
correlation between the black hole mass and the mean optical continuum
luminosity, with the dependence $M \propto L^{0.5}$, suggesting that
more luminous AGN radiate at a larger fraction of the Eddington
luminosity. \citet{nw96} found a correlation between \sigs \ and the
radio luminosity at 1415 MHz, L$_{1415}$, in Seyfert galaxies. This
relation was linked to previously known correlations between radio
power and absolute blue luminosity \citep{meurs84,edelson87,w92c} and
between radio power and [OIII] line width \citep[{\it e.g.} ][]{w85}.
They speculated that the origin of such a correlation might result
from the correlation of \Mbh \ with bulge mass and that the radio
sources associated with larger central engines might be
correspondingly more luminous.  In Figure \ref{fig:bhsigrad} we plot
$L_{1415 MHz}$ vs. \Mbh, plotting different symbols for radio-loud ($R
> 8.0, +$), radio-intermediate($1.0 < R < 8.0, \circ$) and radio-quiet
AGN($R < 1.0, \bullet$). The criteria are chosen rather arbitrarily,
but nevertheless demonstrate some interesting differences. The solid
line shows a fit to the most radio-quiet group.  The correlation is
quite strong ($R=0.82$, P(null)=0.003\%) with a slope $0.84\pm0.12$. A
fit including radio-intermediate data (not shown) is somewhat weaker
($R=0.58$, P(null)=0.1\%) and a fit to the entire sample weaker still
($R=0.40$, P(null)=2\%). The results suggest that at least in
radio-quiet AGN the radio luminosity, emitted on kpc scales is linked
to the mass of the nuclear black hole.  \citet{nw96} noted other
explanations for their result, including higher ambient pressure in
larger bulges leading to increased radio emissivity.  We cannot
distinguish between these possibilities and it is possible that both
effects work together to produce the observed correlation.

\section{Summary}

Using the assumption that NLR kinematics are predominantly due to
virial motion in the host galaxy potential, we find that AGN follow a
relation between \Mbh \ and \sige \ similar to normal
galaxies. Although the AGN black hole masses are determined using
reverberation mapping, we find no significant differences with results
from studies using stellar dynamical techniques. Satisfied that the
black hole masses in AGN are reliable we can begin to examine the
relationship of black hole mass with other properties of AGN.  As an
example, we have found that the radio luminosity in radio-quiet AGN is
correlated with black hole mass relating parameters determined on
vastly different size scales.

I would like to thank Donna Weistrop for her support and comments on a
draft of this letter. This work was supported in part by NASA, under
contract NAS5-31231.

\end{document}